\documentclass[preprint,showpacs,preprintnumbers,amsmath,amssymb]{revtex4}
\usepackage{graphicx,color}

\begin{document}      

\title{Effects of kappa distribution function on Landau damping in electrostatic Vlasov simulation}

\author
{Yue-Hung Chen, Yasutaro Nishimura, and Chio-Zong Cheng}

\address{Institute of Space, Astrophysical, and Plasma Sciences \\
Plasma and Space Science Center \\
National Cheng Kung University, Tainan 70101, Taiwan}

\begin{abstract}
Effects of non-thermal high-energy electrons on Langmuir wave-particle interaction
are investigated by an initial value approach.
A Vlasov-Poisson simulation is employed which
is based on the splitting scheme by Cheng and Knorr
[Cheng, C.Z. and G.~Knorr, 1976: J. Comput. Phys. {\bf 22}, 330-351.].
The kappa distribution function is taken as an example of non-thermal electrons.
The modification is manifested as an increase in the Landau damping rate and 
a decrease in the real frequency for a long wavelength limit.
A part of the analyses by the modified plasma dispersion function 
[Summers, D. and R.M.Thorne, 1991: Phys. Fluids, B {\bf 3}, 1835-1847.]
is reproduced for $\kappa = 2,3$ and $6$. The dispersion relation
from the initial value simulation and the plasma dispersion function
compare favorably.
\end{abstract} 


\maketitle

\section{Introduction}
Space plasma is far from being in a thermal equilibrium.
Suprathermal electrons are often observed in space plasmas (Vasyliun 1968). 
The kappa distribution function (Leubner 2004) is one of the good examples
of non-Maxwellian distribution functions.
In one limit (Tsallis 1988) the kappa distribution function 
evolves toward Maxwellian.

To investigate wave-particle interaction, or Landau damping (Landau 1946; Jackson 1960)
one needs to incorporate 
the velocity space dynamics by the Vlasov equation.
One of the first pieces of work which made the numerical simulation of Vlasov equation 
available is the splitting scheme by Cheng and Knorr (Cheng 1976),
which is based on the method of characteristics.
This method has become a standard method for the Vlasov type simulation.
The method applies as long as the system is dissipation-less, in other words, 
if the phase volume of the system conserves.
The Vlasov-type simulation in lower dimensional cases has advantages over 
Particle-in-Cell (PIC) simulation, since the Vlasov simulation does not accompany statistical errors.
The Vlasov simulation in lower dimension is suitable for investigating subtle 
effects such as a slight deviation of the equilibrium distribution function
from Maxwellian.  

One of our long term goals is to investigate the dynamics of 
Langmuir solitons (Zakharov 1972) which can then evolve into 
Langmuir turbulence (Wang 1994, 1995, and 1996).
By the splitting scheme (Cheng 1976),
the electrostatic Vlasov simulation has revealed 
heating of plasmas by Langmuir solitons (Li 1995).

The purpose of this paper is two-fold.
One purpose is to recapitulate the method of (Cheng 1976) accurately
for the further advanced study of Langmuir solitons.
In this work, the splitting scheme is revisited and the results 
of (Cheng 1976) are reproduced.
The other purpose is, as an initial exercise, to capture the effect of
non-Maxwellian distribution function on Landau damping. 

This paper is organized as follows.
In Sec.\ref{s2}, the basic computation model is described.
In Sec.\ref{s3}, we then start from verifying the simulation results 
with free streaming case whose analytical solutions are known.
The benchmark linear and nonlinear numerical simulation results are discussed in Sec.\ref{s4}. 
The effects of non-Maxwellian kappa distribution function are discussed in Sec.~\ref{s5}. 
Direct comparison of simulation results with modified plasma dispersion function
is discussed in Sec.~\ref{s6}.
We summarize this work in Sec.~\ref{s7}.

\section{Model equations and numerical methods}
\label{s2}
In this section, the model equation of the Vlasov-Poisson simulation is described.
For the transparency of the work,
we recapitulate Cheng (1976) as precise as possible including the notations.
A one-dimensional Vlasov-Poisson system in the MKS unit is given by 
\begin{equation}
\frac{\partial  }{\partial t} f_j   + v \frac{\partial }{\partial x} f_j  + 
\frac{q_j E}{ m_j} \frac{\partial }{\partial v} f_j   = 0 ,
\end{equation}
and
\begin{equation}
\frac{\partial E}{\partial x} = \frac{e}{\varepsilon_0 } \left( n_i  - n_e \right) ,
\end{equation}
where the densities $n_j = n_j (x,t)$ of each species 
(subscripts $j=i$ for the ions and $j=e$ for the electrons)
are given by the distribution functions $f_j = f_j (x,v,t)$
\begin{equation}
n_j (x,t)=  \int_{-\infty}^\infty f_j (x,v,t) dv .
\end{equation}
Here $m_j$ and $q_j$ are the mass and the charge of the species.
The electric field is given by $E$ and the vacuum permittivity is given by $ \varepsilon_0$.
The coordinates in configuration and velocity spaces are given 
by $x$ and $v$, respectively.

The equations are further normalized by the Debye length
$\lambda_e = \left( \varepsilon_0 T_e / n_0 m_e \right)^{1/2}$, 
the plasma frequency $\omega_e = \left( n_0 e^2 / \varepsilon_0 m_e \right)^{1/2}$,
and the electrostatic field is normalized by
$\bar{E} = (e \lambda_e E /T_e ) $ which is equivalent to having 
the electrostatic energy being comparable to the electron thermal energy.
Here, $e$ is the unit charge and the electron temperature is given by $T_e$.
By employing the bars denoting the normalized values, 
we have $\bar{\bf x} = {\bf x} / \lambda_e , \bar{t} =  \omega_e t, 
\bar{\bf v} = {\bf v} / \lambda_e \omega_e $, and 
$\bar{f} = ( \lambda_e \omega_e / n_0 ) f  $, where 
$n_0$ is the equilibrium electron and ion densities.  
Note that the thermal velocity of the electrons are
given by $v_e = (T_e/m_e)^{1/2} = \lambda_e \omega_e$.

After the normalization, we obtain a Vlasov-Poisson system
in the $\lambda_e  $ (space) and $\omega_e  $ (time) scales,
\begin{equation}
\frac{\partial}{\partial \bar t} {\bar f} 
+ {\bar v} \frac{\partial}{\partial \bar x} {\bar f} ({\bar x},{\bar v},{\bar t}) 
- {\bar E} \frac{\partial}{\partial \bar v}  {\bar f} ({\bar x},{\bar v},{\bar t}) = 0 ,
\label{nvlasov} 
\end{equation}
and
\begin{equation}
\frac{\partial {\bar E}  }{\partial \bar x}  
= 1 - \int_{-\infty}^\infty {\bar f} ({\bar x},{\bar v},{\bar t}) {\bar {dv}} ,
\label{npoisson} 
\end{equation}
where the ion density is taken to be uniform [signified by unity in Eq.(\ref{npoisson})]
and the distribution function ${\bar f}$ is for the electrons.
Equations (4) and (5) correspond to Eqs.(1a) and (1b) of Cheng (1976).
Hereafter we drop the bars.

The splitting scheme (Cheng 1976) is based on the method of characteristics 
which is equivalent to the lowest order symplectic integrator (Ruth 1983).
We evolve the distribution function by tracing the characteristic curves
in the phase space. The method takes three steps which is given by
\begin{equation}
f^{*} \left( x, v \right) = f^{n} \left( x - v \Delta t /2 , v \right),
\label{sp1} 
\end{equation}
\begin{equation}
f^{**} \left( x, v \right) = f^{*} \left( x, v + E (x) \Delta t \right),
\label{sp2} 
\end{equation}
and finally,
\begin{equation}
f^{n+1} \left( x, v \right) = f^{**} \left(x - v \Delta t /2, v \right),
\label{sp3} 
\end{equation}
whose kinetic energy part and the potential energy part are time
advanced alternatively (the lowest order method corresponds
to the well-known leap frog method employed frequently in PIC simulation).
The superscript $n$ stands for the time step.

Note that the splitting scheme is not a finite difference scheme. 
In the splitting scheme, if the reference points along the characteristic curves
``$x - v \Delta t /2$'' or ``$v + E (x) \Delta t $''
are exactly on the mesh points, the method is quite trivial.
However, in general, the points of references are located in between 
mesh points of the $x$ and $v$ space.
We thus need an interpolation technique to realize Eqs.(\ref{sp1}),(\ref{sp2}), and (\ref{sp3}). 
As in Cheng (1976), we use Fourier interpolation in the configuration space and
linear interpolation in the velocity space (Watanabe 2005).
The computational mesh we employed is exactly that of Fig.1 in Cheng (1976).
Note that we do not have mesh points on the $v=0$ axis.
 
To make the simulation self-consistent we need to solve Poisson equation, Eq.(\ref{npoisson}).
The Poisson equation is solved by the Fourier transform since we adopt
a periodic configuration in $x$ in this paper.
All the calculations in this paper employ periodic boundary conditions.
The electric field is solved directly as in Eq.(\ref{npoisson}) without calculating
the electrostatic potential.

\section{Free streaming case with $E=0$}
\label{s3}
\indent
To begin with the numerical simulation, we start by verifying the solution
of free streaming case [the Van Kampen mode, (Van Kampen 1955)] whose analytical solutions are known.
Setting $E=0$ in Eq.(\ref{nvlasov}), we obtain
\begin{equation}
\frac{\partial}{\partial t} f (x,v,t) + \frac{\partial}{\partial x} f (x,v,t) = 0.
\end{equation}
Following Nicholson 1992, for example, the analytical solution can be given by setting
\begin{equation}
f(x,v,t) \sim \exp{(-i k v t)} .
\end{equation}
For example, if we take an initial condition
\begin{equation}
f(x,v,0) = A \cos{ ( k x)} e^{-v^2/2} 
\label{ic1} 
\end{equation}
we obtain the analytical solution
\begin{equation}
f(x,v,t) = A \cos{ ( k x)} e^{-v^2/2} \cos{(k v t)}.
\end{equation}

Shown in Fig.1(a) is the solution of Vlasov equation in a free-streaming case.
No force is acting on the electrons.
The thick solid line is the initial condition of the distribution function
at a fixed point $x=0$. The time advanced distribution function by
the numerical simulation is given by the black dots at $t=12.5$ 
(time is normalized by the inverse of plasma frequency, $\omega_{e}^{-1}$),
which matches with the analytical solution given by the dash-dotted curves.
In the simulation, parameters employed are the maximum cut-off velocity
$v_{max}=10  v_{e}$, and $0 \le x \le 4 \pi$.
For the mesh points, 32 and 256 are taken in the $x$ and $v$ direction, 
respectively (although we did calculate, the regions $|v|>5.0$ are not shown in the figure).
Shown in Fig.1(b) is the solution of Vlasov equation in the free-streaming case
but with an initial condition given by
\begin{equation}
f(x,v,0) = [ 1 + A \cos{ ( k x)} ] e^{-v^2/2} .
\label{ic2} 
\end{equation}
The time advanced distribution function at $t=6.25$  is given by the solid curve at 
$x=\pi$ and by the dash-dotted curve at $x= 3 \pi $. 
Note that, as demonstrated by the two solutions at $x=\pi$ and   $x= 3 \pi $,
the evolution of the distribution function exhibits point reflection across $x=2 \pi$.
The distribution function streams in positive direction in 
$v>0$ while streams in negative direction in $v<0$, as a reminder.
In both Fig.1(a) and Fig.1(b), $A=0.5$ and $k=0.5$ are taken.
The calculation discussed in this section validates the interpolation
scheme employed for the Vlasov equation.

\section{Benchmark of linear and nonlinear simulation results}
\label{s4}
\indent
In this section, linear and nonlinear simulation results are
compared and benchmarked with those of Cheng (1976).
For both the linear and the nonlinear simulation, we take an initial condition of 
the form of Eq.(\ref{ic2}). 

In the linear simulation, parameters employed are exactly those of 
Cheng (1976): the maximum cut-off velocity
$v_{max}=4.0 v_{e}$ and $0 \le x  \le 4 \pi$,
and the mesh points are 8 and 32 for $x$ and $v$, respectively.  
In Fig.2, the Landau damping phase is shown in terms of electric field strength $|E|$. 
As in Cheng (1976), recurrence effect takes place after $t = 42$.
Note that only the Fourier component $k=0.5$ is kept and the other modes are filtered out in the linear calculation.
A mesh point $x= \pi$ is chosen for a diagnostic point in Fig.2.

The figure corresponds to Fig.3 of Cheng (1976) except that $A=0.01$
instead of $A=0.5$ is taken. The measured frequency and the damping rates
are $\omega=1.41$ and $\gamma = -0.155$, respectively.

The nonlinear simulation is shown in Fig.3.
In the simulation of Fig.3, $A=0.5$ and $k=0.5$ are taken.
Figure 3(a) shows the time evolution of electric field at a fixed point $x=\pi$.
Instead of monotonic decrease the saturation of the amplitude can be seen after $t=20$.  
Figure 3(b) shows the distribution function at $t=75$ at a fixed point $x=\pi$. 
In Fig(b), we can see a local flattening of the distribution function
in the vicinity of the phase velocity. The phase velocity of the Langmuir wave estimated
by the linear theory is $\omega/k = 2.82$.

Note that in the nonlinear simulation, as suggested in Cheng (1976), 
the frequencies of all the higher modes come into play at the later stage.
As a result, resonance occurs at multiple locations in the velocity space
and thus microscopic structures are generated [manifested as {\it wrinkles} in Cheng (1976)]
whose size can be comparable to the mesh size in the velocity space.
To resolve all the resonance, one needs to employ an extremely high resolution
in the velocity space.

\section{Effects of kappa distribution functions}
\label{s5}
\indent
In this section, we investigate the effects of high energy electrons
by employing kappa distribution functions (Leubner 2004) instead of a Maxwellian  
(for the initial condition). 
A kappa distribution function we employed is given by
\begin{equation}
f_v  \left( v \right) \propto \left[ 1 + \frac{v^2}{2 \kappa } \right]^{-\kappa - 1}  .
\label{ic3} 
\end{equation}
Note that Maxwellian and kappa distribution functions are related
\begin{equation}
\lim_{\kappa \rightarrow \infty} f_v (v)  \propto \exp{(-v^2/2 )}.
\label{ic4} 
\end{equation}
The spatial distribution is given in the form of Eq.(\ref{ic2}) for
the initial condition, thus $f(x,v,0) \sim [ 1 + A \cos{ ( k x)} ] f_v (v)$ is given.
We have normalized the kappa distribution function to satisfy $\int_{-\infty}^{\infty} dv f_v(v)=1$,
so that the effective number of electrons will be the same as in the Maxwellian case.
Some of the notable features of the kappa distribution function are shown in Fig.4.
In Fig.4(a), Maxwellian (black) and kappa distributions with $\kappa = 1$
(red) are compared. One can see large population of high-energy tail 
in the $\kappa = 1$ case.
The functions are plotted in the logarithmic scales for $\kappa=1$ (red) , $\kappa=2$ (green),
$\kappa=5$ (blue), and Maxwellian (black) cases in Fig.4(b).

Figure 5 shows the linear damping with cases when $\kappa=1$ (red), $\kappa=2$ (green),
and Maxwellian (black) are taken as initial distribution functions.  
Parameters employed are  
$v_{max}=10 v_{e}$ and $0 \le x \le 4 \pi$ with 32 and 256 mesh points in $x$ and $v$.
As in Fig.2, we have taken $A=0.01$ and $k=0.5$.

With the kappa distribution function 
the Landau damping rate increases and the real frequency decreases
(smaller $\kappa$ has larger effects).
The variation of the linear damping rate $\gamma$ and the real frequencies $\omega$
are summarized in Fig.6(a) and Fig.6(b) as a function of $\kappa$.
The value $\kappa$ varies from $1.0$ to $10.0$.
The values plotted in both Fig.6(a) and Fig.6(b) are normalized by that of Maxwellian
(thus in the figures, $\gamma / |\gamma_{Max}| = -1$ and $\omega / \omega_{Max} = 1$ 
for a Maxwellian initial condition).

From the linear theory (Landau 1946; Jackson 1960)
the Landau damping rate is proportional to the slope of the distribution function
at the phase velocity of the wave [see, for example, Nicholson (1992)]. 
Employing the measured real frequencies $\omega$ of Fig.6(b) (and $k=0.5$ for the wave number), 
the slope of the distribution function $\partial_v f_v(v) $ at the phase velocity $\omega/k$
is estimated and shown in Fig.7. Smaller $\kappa$ cases have more negative values (and thus
larger damping rate) which supports the nature of Fig.6(a).
For small values of $k$, these latter trends (damping rate increasing and the real frequency 
decreasing) are consistent with the analytical work (Chateau 1991; Summers 1991; Thorne 1991).
In the next section, we conduct a direct comparison of the Vlasov simulation
with the roots of the plasma dispersion relation, 
by taking exactly the same distribution functions employed in Summers (1991) and Thorne (1991). 

\section{Direct comparison with modified plasma dispersion function}
\label{s6}
By an analogy of plasma dispersion function (Fried 1961) for Maxwellian,
Summers and Thorne (Summers 1991; Thorne 1991) have extended their work to 
kappa distribution function.
We compare the damping rate and real frequencies
in our simulation with their theoretical work for different values of $\kappa$
and different wave-numbers $k$.

To see the match between the two, in this section we take exactly the same distribution 
function employed in Summers and Thorne (Summers 1991; Thorne 1991).
We have taken initial distribution function in the form (see Appendix)
\begin{equation}
f_v \left( v \right) =
\frac{1}{\sqrt{\pi}} \frac{\sqrt{\kappa}}{\sqrt{2 \kappa -3}}
\frac{\Gamma (\kappa)}{\Gamma ( \kappa -1/2) \kappa^{3/2}}
\left( 1 + \frac{{v}^2}{2 \kappa -3 } \right)^{-\kappa}.
\label{ic5} 
\end{equation}
Note the difference in the exponent part 
(``$-\kappa$'' dependence instead of ``$-\kappa -1 $'') 
between Eq.(\ref{ic5}) and Eq.(\ref{ic3}). 

In Fig.8, we plot the simulation results and the numerical roots of the dispersion relation
for $\kappa=2,3$, and $6$
[reproduced from the modified dispersion function of Summers (1991) and Thorne (1991)].
Figure 8(a) is for the damping rates $\gamma$ versus the wave-numbers $k$.
Figure 8(b) is for the real frequencies $\omega$ versus $k$. 
The roots from the (modified) dispersion relation are plotted as dash-dotted curves. 
The black, red, and green dash-dotted curves are for
$\kappa=2,3$, and $6$, respectively. 

Those obtained from linear numerical simulation are plotted as black circles. 
The damping rate and the real frequencies are obtained from oscillation signal 
of the electric field at a fixed point $x=\pi$.
In this section we did the survey only up to $k=2.0$. 
When the magnitudes of linear damping rate and the real frequencies become comparable,
measurement of $\gamma$ and $\omega$ becomes troublesome.
The dispersion relation from the initial value simulation compare favorably
with that from the plasma dispersion function.
With the smaller $\kappa$ values, the real frequency decreases.
Note that, however, with the smaller $\kappa$ values, the absolute value of
the damping rates can also decrease for larger values of $k$, 
contrary to what we have obtained in the previous section. 
The damping rates are sensitive function of the local value $\partial_v f$ 
where the resonant phase velocities ``$\omega /k$'' are located at,
and can vary depending on the wave-numbers $k$.

One of the advantages of the initial value approach is its application
to nonlinear simulation.
Our preliminary nonlinear simulation results employing a kappa distribution
function as an initial condition are presented below.
Figure 9 shows electron distribution functions suggesting long time evolution up to $t=1500$.
The distribution functions are given at a fixed point $x=\pi$.
Figure 9(a) is for a Maxwellian and
Fig.9(b) is for a kappa distribution function ($\kappa=2$). 
The dash-dotted curves are for $t=0$ and the solid curves are for $t=1500$.
The integration of the distribution functions over the
velocity space is the same for the two cases.
The integration of the ion density over the configuration space
is kept the same with the electron density (total numbers of ions and electrons in the system are the same).
In Fig.9, a relatively large cut-off velocity 
$v/v_{the}=12.0$ is taken with a high resolution (1024 mesh points
are taken in the velocity space). 
The $\kappa=2$ distribution function have larger population at the high energy tail.
In the simulation of Fig.9, $A=0.5$ and $k=0.5$ are taken.
Figure 9(c) shows local expansion of Fig.9(a) and Fig.9(b) near the resonant phase velocities.
In the figure, the perpendicular lines are suggesting the phase velocities.
Note that the dash-dotted black line tends to solid black line  
(frequency down-shift for Maxwellian)
while the dash-dotted red line tends to solid red line
(frequency up-shift for kappa function).
In a very long time scale, normal mode frequency changes 
and the distribution functions can possibly evolve toward a similar equilibrium state.

\section{Summary}
\label{s7}
In this work, the splitting scheme is revisited and the simulation results 
are compared with Cheng (1976). Based on the validation of the code,
as an initial exercise, we have discussed the effect of 
non-Maxwellian distribution function by employing the kappa-distribution function.

The slope [the absolute value of $\partial_v f(v)$] of the distribution function at the phase velocity is estimated
which supports the nature of increasing damping rate at smaller $\kappa$ values.
The simulation results compare favorably with the analyses based on
modified plasma dispersion function (Summers 1991; Thorne 1991).
The specific calculations we have demonstrated are for $\kappa=2,3$, and $6$, with
the wave-numbers $k=0.5, 1.0$ and $2.0$.
Our preliminary nonlinear simulation employing the kappa distribution
function is presented.

A part of this work is supported by National Cheng Kung University
Top University Project and a part by National Science Council of Taiwan,
NSC 100-2112-M-006-021-MY3.
One of the authors YN would like to thank Professor Yasushi Nishida for discussions.
The authors thank one of the referees for suggesting several pieces of work on the analysis 
of kappa distribution function.

\appendix
\section{A brief review of modified plasma dispersion function}
\label{a1}
\indent
We review the modified plasma dispersion function 
(the ``$Z_{\kappa}^{*}$ function'') for kappa distribution functions (Summers 1991). 
In this appendix, we invert the notation (see Sec.\ref{s2}); 
the values with bars are normalized ones, and all other values are those before normalization. 
Neglecting the ion contribution and requiring 
the longitudinal component of the dielectric tensor to be zero (Fried 1961), we obtain the
Langmuir wave dispersion relation
\begin{equation}
1 + \frac{\omega_e^2}{n_0 k^2} 
\int_{-\infty}^{\infty} dv \frac{k \partial_v f_v}{\omega - k v} = 0.
\label{ap1} 
\end{equation}
Here, we have assumed a planer wave of the form $\exp{ ( i k x - i \omega t )}$
for all the perturbed quantities, where $i$ is the imaginary unit.
Equation (\ref{ap1}) is exactly what we have in our Vlasov system employed for the numerical simulation.
For a one dimensional Maxwellian
\begin{equation}
f_v  \left( f \right) = \frac{n_0}{\sqrt{\pi} v_t} \exp{\left(- v^2 /v_t^2 \right)},
\label{ap2}
\end{equation}
by introducing the plasma dispersion function (the ``$Z$ function'', Fried 1961)
\begin{equation}
Z \left( \xi \right) = \frac{1}{\sqrt{\pi}} \int_{- \infty}^{\infty} \frac{e^{-s^2}}{s - \xi} ds ,
\label{ap3} 
\end{equation}
we arrive at the well-known Langmuir wave dispersion relation
\begin{equation}
1 + \frac{2 \omega_e^2}{k^2 v_t^2} \left[ 1 +  \frac{\omega}{k v_t} Z \left( \frac{\omega}{k v_t} \right) \right] = 0 ,
\label{ap4} 
\end{equation}
where $v_t^2 = 2 T_e / m_e = 2 v_e^2$.

In contrast, Summers (1991) employs a one dimensional kappa distribution function
\begin{equation}
f_v  \left( v \right) = \frac{n_0}{\sqrt{\pi} v_e} 
\frac{\sqrt{\kappa}}{\sqrt{2 \kappa -3}}
\frac{\Gamma (\kappa)}{\Gamma ( \kappa -1/2) \kappa^{3/2}}
\left( 1 + \frac{{v}^2}{2 \kappa -3 } \right)^{-\kappa}.
\label{ap5}
\end{equation}
Here, $\Gamma (\kappa) = \int_{0}^{\infty} e^{-s} t^{\kappa-1} ds$ stands for Gamma function.
Substituting Eq.(\ref{ap5}) into Eq.(\ref{ap1}) 
and by introducing a {\it modified} plasma dispersion function,
\begin{equation}
Z_{\kappa}^{*} \left( \xi \right) 
= \frac{1}{\sqrt{\pi}}  
\frac{\Gamma (\kappa)}{\Gamma ( \kappa -1/2) \kappa^{3/2}}
\int_{- \infty}^{\infty} \frac{ \left( 1 + s^2/ \kappa \right)^{-\kappa-1} }{s - \xi} ds ,
\label{ap6} 
\end{equation}
we arrive at the dispersion relation for kappa distribution function
\begin{equation}
1 + \frac{2 \omega_e^2}{k^2 v_e^2} \frac{\kappa }{2 \kappa - 3} 
\left[ 1  - \frac{1}{2 \kappa} +  \xi
Z_{\kappa}^{*} \left( \xi \right) \right] = 0
\label{ap7} 
\end{equation}
where
\begin{eqnarray*}
\xi = \frac{\sqrt{\kappa}}{\sqrt{2 \kappa -3}} \frac{\omega}{k v_e} .
\end{eqnarray*}
Note that the exponent ``$- \kappa -1$'' in Eq.(\ref{ap6}) arises from taking a derivative 
on the right hand of Eq.(\ref{ap5}). 
A normalized dispersion relation is given by
\begin{equation}
\xi Z_{\kappa}^{*} (\xi)  + 1 -\frac{1}{2 \kappa} + \frac{ \kappa - 3/2}{\kappa} \bar{k}^2=0
\label{ap8} 
\end{equation}
where $\bar{k} = k \lambda_e$.

Following Thorne (1991), the root finding algorithm employed in Fig.8 is stated.
We first let $\xi = x + i y$, where $x$ and $y$ are real numbers.
We then fix the $y$ value and solve the imaginary part of the dispersion
relation Eq.(\ref{ap8}), $ Im \left[ \xi Z_{\kappa}^{*} (\xi)  \right] =0$, to obtain $x$.
When both $x$ and $y$ are given, we solve the real part of Eq.(\ref{ap8}) for $\bar{k}$ by
\begin{eqnarray*}
\bar{k}=\sqrt{\frac{\kappa}{\kappa-3/2}}\sqrt{ \frac{1}{2 \kappa} -1- Re \left[ \xi  Z_{\kappa}^{*} (\xi)  \right] }.
\end{eqnarray*}
Finally, the real frequencies and the damping rates are given by
$\bar{\omega}= \bar{k} x \sqrt{(2 \kappa-3)/\kappa}$ and
$\bar{\gamma}= - \bar{k} y \sqrt{(2 \kappa-3)/\kappa}$. 
As in Sec.\ref{s2}, the time scale is normalized by $\omega_e^{-1}$.
The black, red, and green curves in Fig.8 are obtained by this latter algorithm.

\newpage
{\bf{References}}

Chateau, Y.F. and N. Meyer-Vernet, 1991:
Electrostatic noise in non Maxwellian plasmas: generic properties and Kappa distributions.
J. Geophys. Res. {\bf 96},5825–5836. 
doi:10.1029/90JA02565. 

Cheng, C.Z. and G.~Knorr, 1976: 
Integration of Vlasov equation in configuration space.
J. Comput. Phys. {\bf 22}, 330-351.
doi:10.1016/0021-9991(76)90053-X.

Fried, B.D. and S.D.Conte, 1961: 
{\it The Plasma Dispersion Function}, 
Academic Press, 1 pp.

Jackson, J.D., 1960: 
Longitudinal plasma oscillations.
J. Nuclear Energy, Part C {\bf 1}, 171-189.
doi:10.1088/0368-3281/1/4/301.

Landau, L. D., 1946: 
On the vibration of the electronic plasma.
J. Phys. (U.S.S.R.) {\bf 10}, 25.

Leubner, M.P., 2004:  
Fundamental issues on kappa-distributions in space plasmas and interplanetary proton distributions. 
Phys. Plasmas {\bf 11}, 1308-1316,
doi:10.1063/1.1667501. 

Li, C.H., J.K.Chao, and C.Z.Cheng, 1995: 
One‐dimensional Vlasov simulations of Langmuir solitons.
Phys. Plasmas {\bf 2}, 4195-4203. 
doi:10.1063/1.871045.

Nicholson, D.R., 1992: 
{\it Introduction to Plasma Theory 2nd ed.}, 
Krieger Publishing, 82 pp. and 120 pp.

Ruth, R.D., 1983: 
A canonical integration technique.
IEEE Transactions on Nuclear Science {\bf 30}, 2669-2671,
doi:10.1109/TNS.1983.4332919.

Summers, D. and R.M.Thorne, 1991: 
The modified plasma dispersion function.
Phys. Fluids, B {\bf 3}, 1835-1847,
doi:10.1063/1.859653.
	
Thorne, R.M. and D. Summers, 1991: 
Landau damping in space plasmas.
Phys. Fluids, B {\bf 3}, 2117-2123,
doi:10.1063/1.859624.

Tsallis, C., 1988: 
Possible generalization of Boltzmann-Gibbs statistics.
J. Stat. Phys. {\bf 52}, 479-487,
doi:10.1007/BF01016429.

Van Kampen, N.G., 1955: 
On the theory of stationary waves in plasmas.
Physica {\bf 21}, 949-943,
doi:10.1016/S0031-8914(55)93068-8.

Vasyliun, V.M., 1968:
A survey of low-energy electrons in evening sector of magnetosphere with OGO 1 and OGO 3.
J. Geophys. Res. {\bf 73}, 2839-2884,
doi:10.1029/JA073i009p02839.

Wang, J.G., G.~L.~Pain, D.~F.~Dubois, and H.~A.~Rose, 1994: 
One-dimensional simulations of Langmuir collapse in a radiation-driven plasma.
Phys. Plasmas {\bf 1}, 2531-2546, 
doi:10.1063/1.870581.

Wang, J.G., G.~L.~Pain, D.~F.~Dubois, and H.~A.~Rose, 1995: 
Vlasov simulation of modulational instability and Langmuir collapse.
Phys. Plasmas {\bf 2}, 1129-1139,
doi:10.1063/1.871391.

Wang, J.G., G.~L.~Pain, D.~F.~Dubois, and H.~A.~Rose, 1996: 
Comparison of Zakharov simulation and open boundary Vlasov simulation of strong Langmuir turbulence.
Phys. Plasmas {\bf 3}, 111-121, 
doi:10.1063/1.871837.

Watanabe, T.H., 2005: 
Vlasov Simulation of the Microturbulence.
J. Plasma and Fusion Research {\bf 81}, 686-697.

Zakharov, V.E., 1972: 
Collapse of Langmuir waves.
Sov. Phys. JETP {\bf 35}, 908-914.

\begin{figure}[ht]
 \includegraphics[height=10cm]{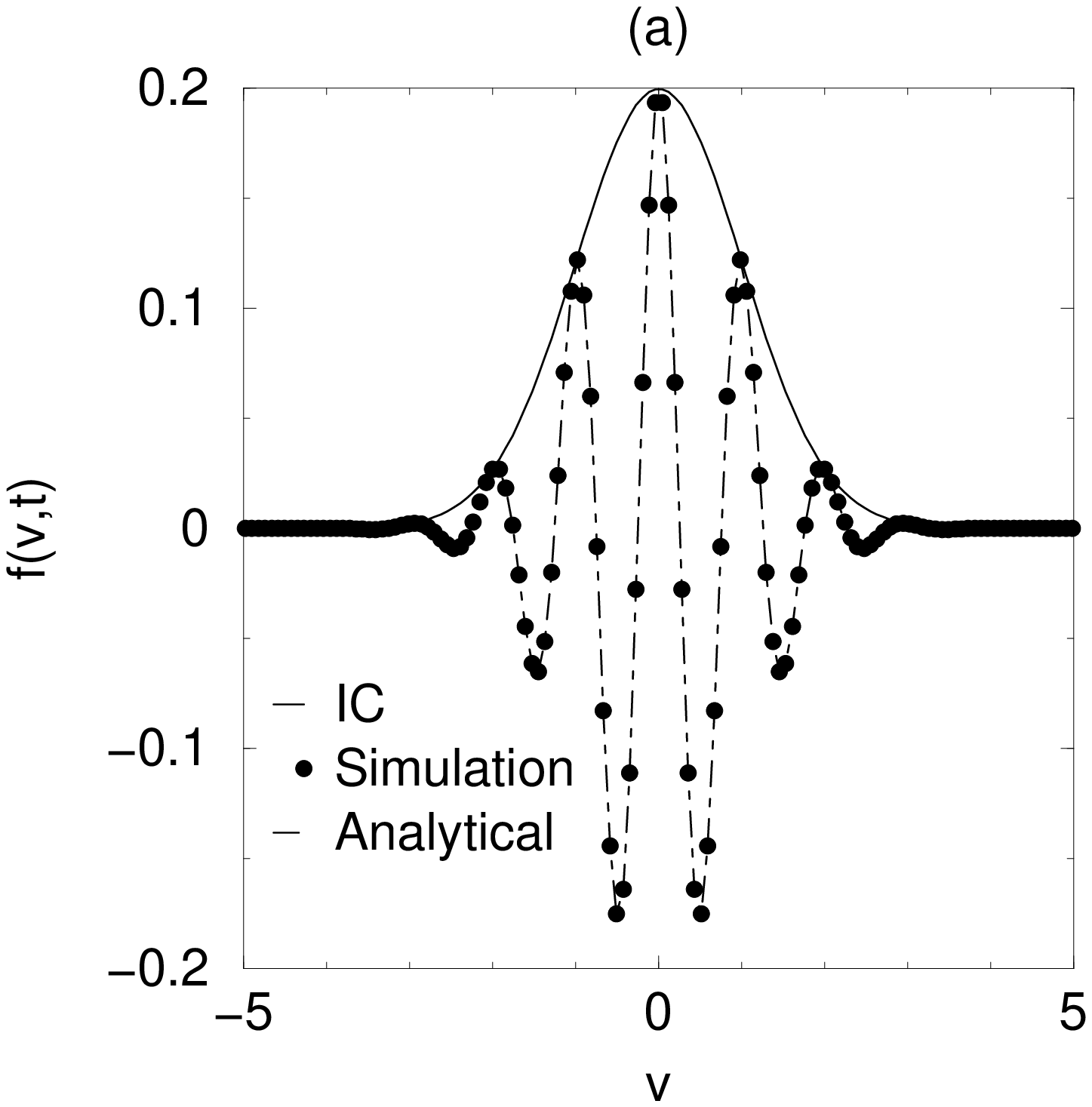}
 \includegraphics[height=10cm]{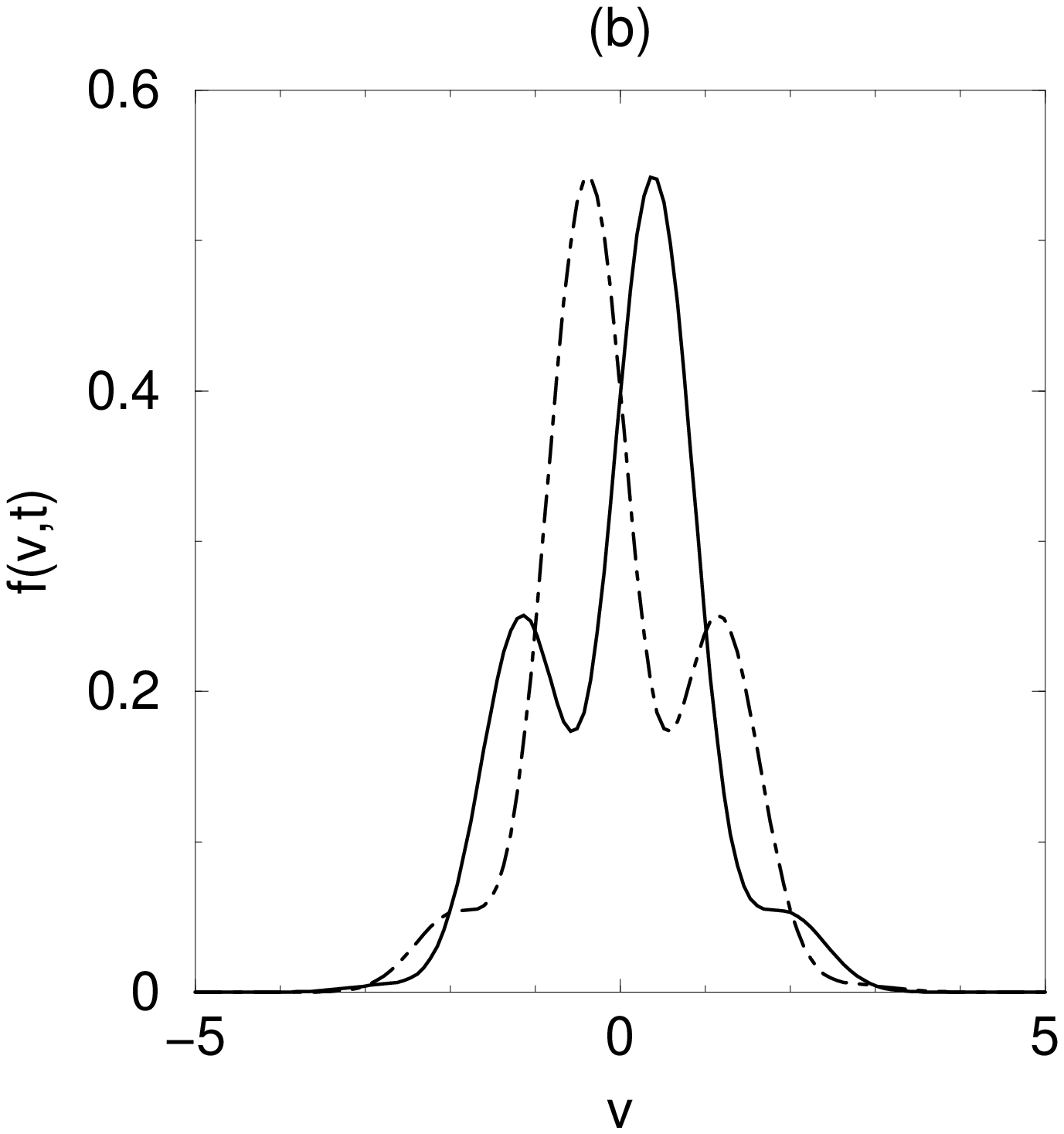}
\vspace{-0.5cm}
\caption{Evolution of distribution functions in free streaming cases.
(a) Distribution functions at a cross section $x=0$ when Eq.(\ref{ic1}) is taken 
as an initial condition.
The thick solid line represents the initial distribution function.
The time advanced distribution function is given by
the black dots, while the analytical solution is given by the dash-dotted curve.
(b) Time advanced distribution function at $x=\pi$ (solid curve) and $x=3 \pi$ (dash-dotted curve)
when Eq.(\ref{ic2}) is taken as an initial condition. }
\label{fig1}
\end{figure}

\newpage
\begin{figure}[ht]
  \includegraphics[height=10cm]{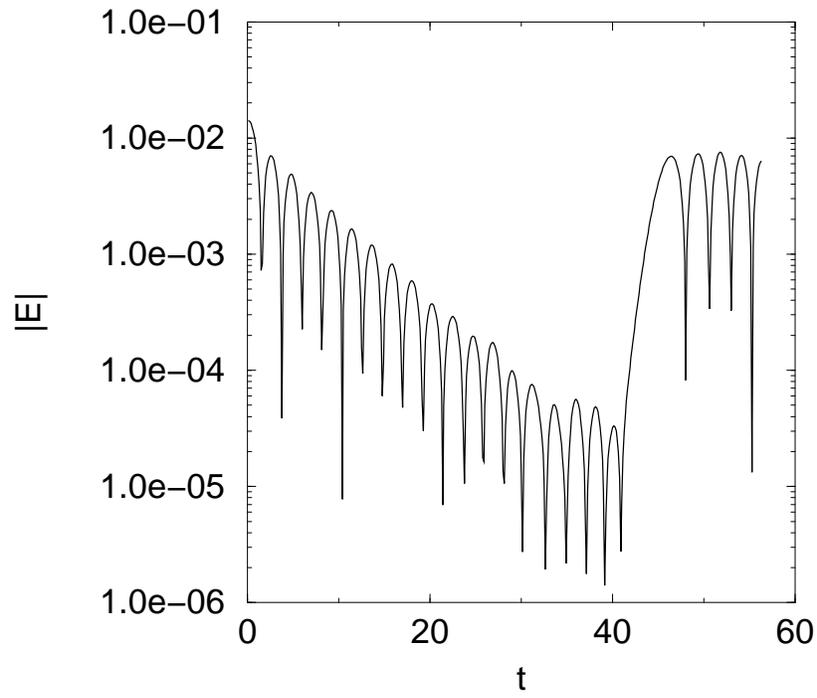}
\caption{Linear simulation results.
Damping of the electric field strength $|E|$ is shown.
Here, $v_{max}=4.0 v_{e}$, and $0 \le x  \le 4 \pi$.
The mesh points are 8 and 32 for $x$ and $v$, respectively.  }
\label{fig2}
\end{figure}

\newpage
\begin{figure}[ht]
  \includegraphics[height=10cm]{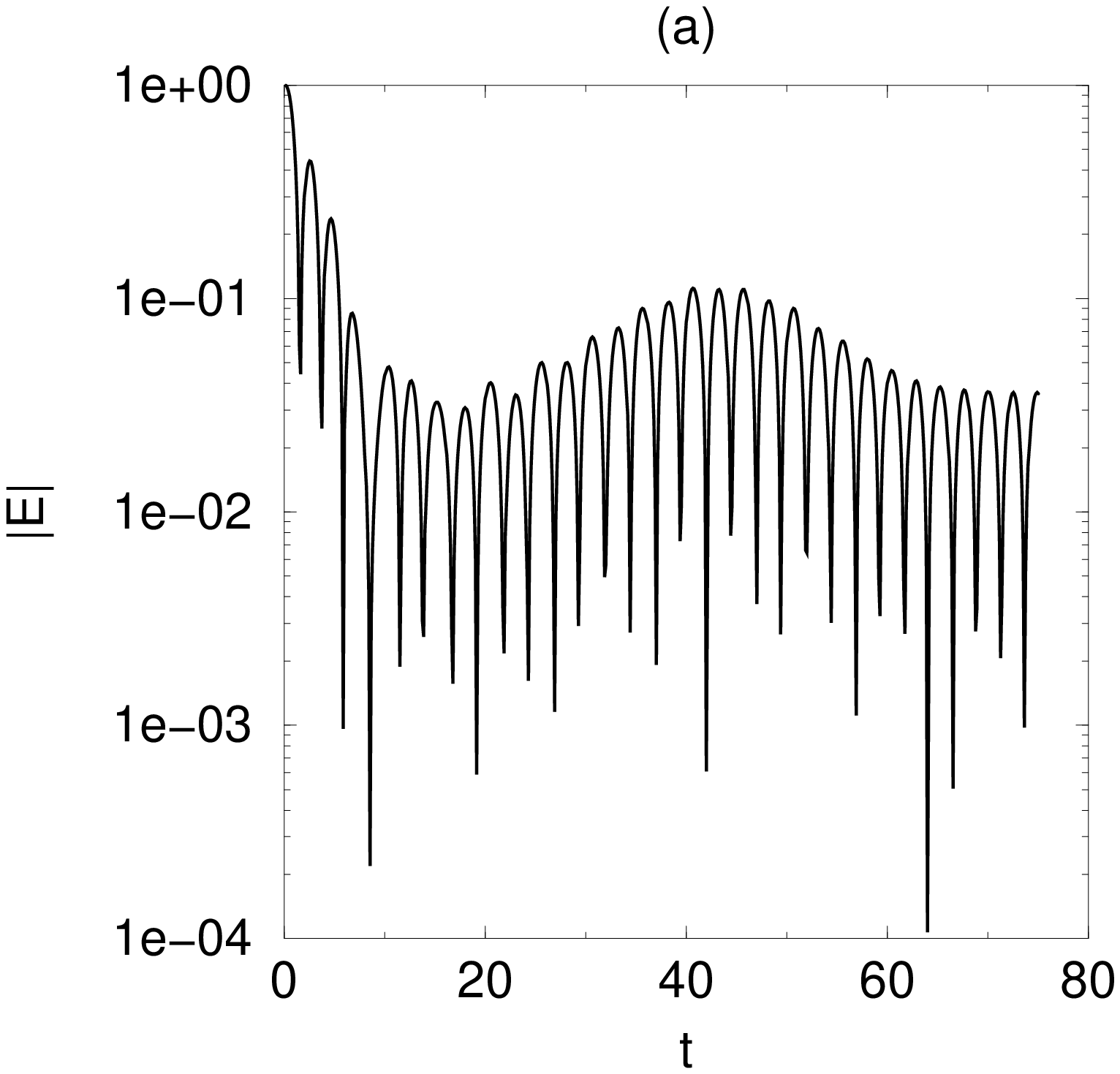}
  \includegraphics[height=10cm]{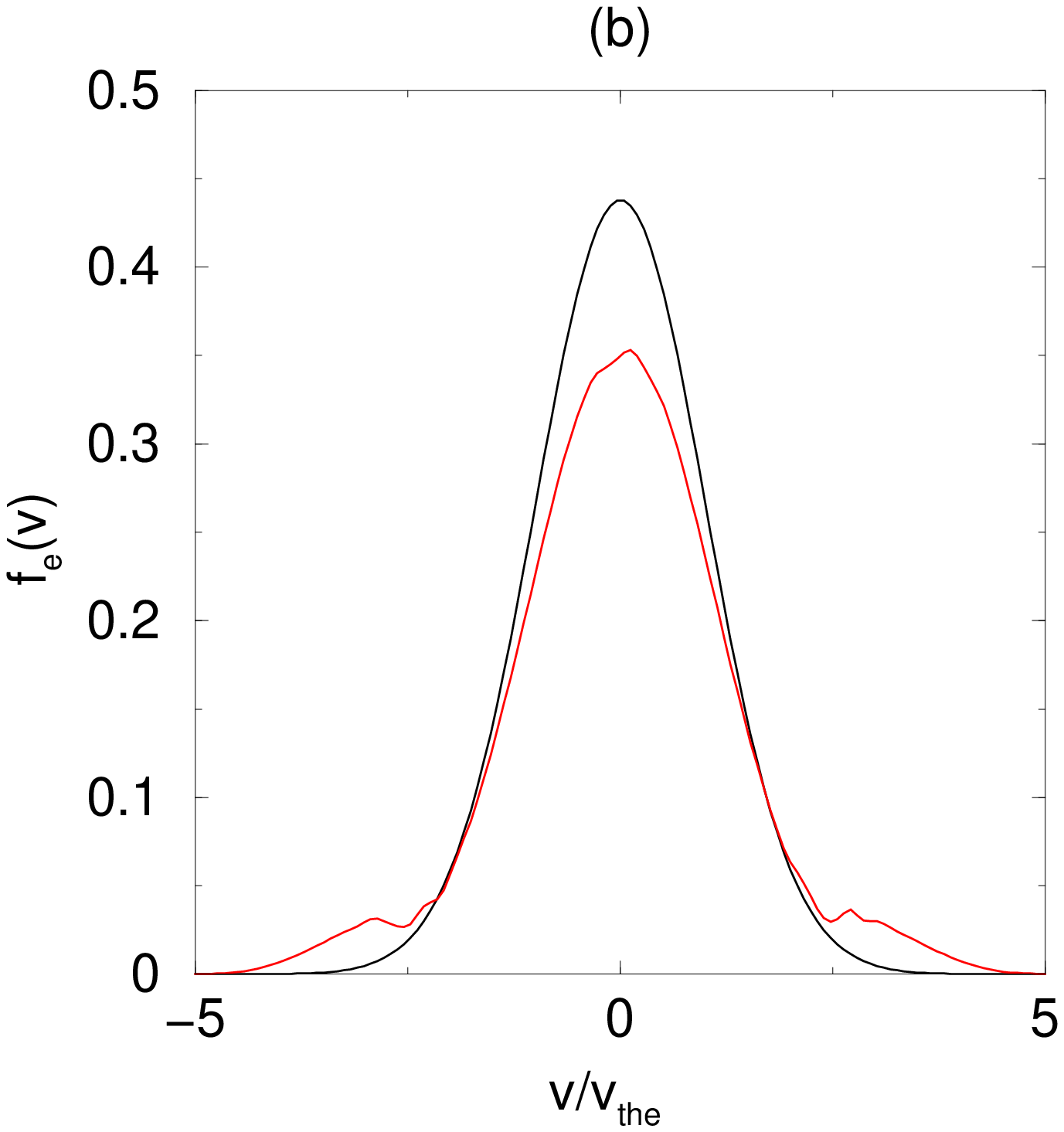}
\caption{Nonlinear simulation results: 
(a) Time evolution of electric field strength $|E|$; 
(b) The initial distribution function (black solid curve) and the
distribution function at $t=75$ (red solid curve), 
both given at $x=\pi$.}
\label{fig3}
\end{figure}

\newpage
\begin{figure}[ht]
  \includegraphics[height=10cm]{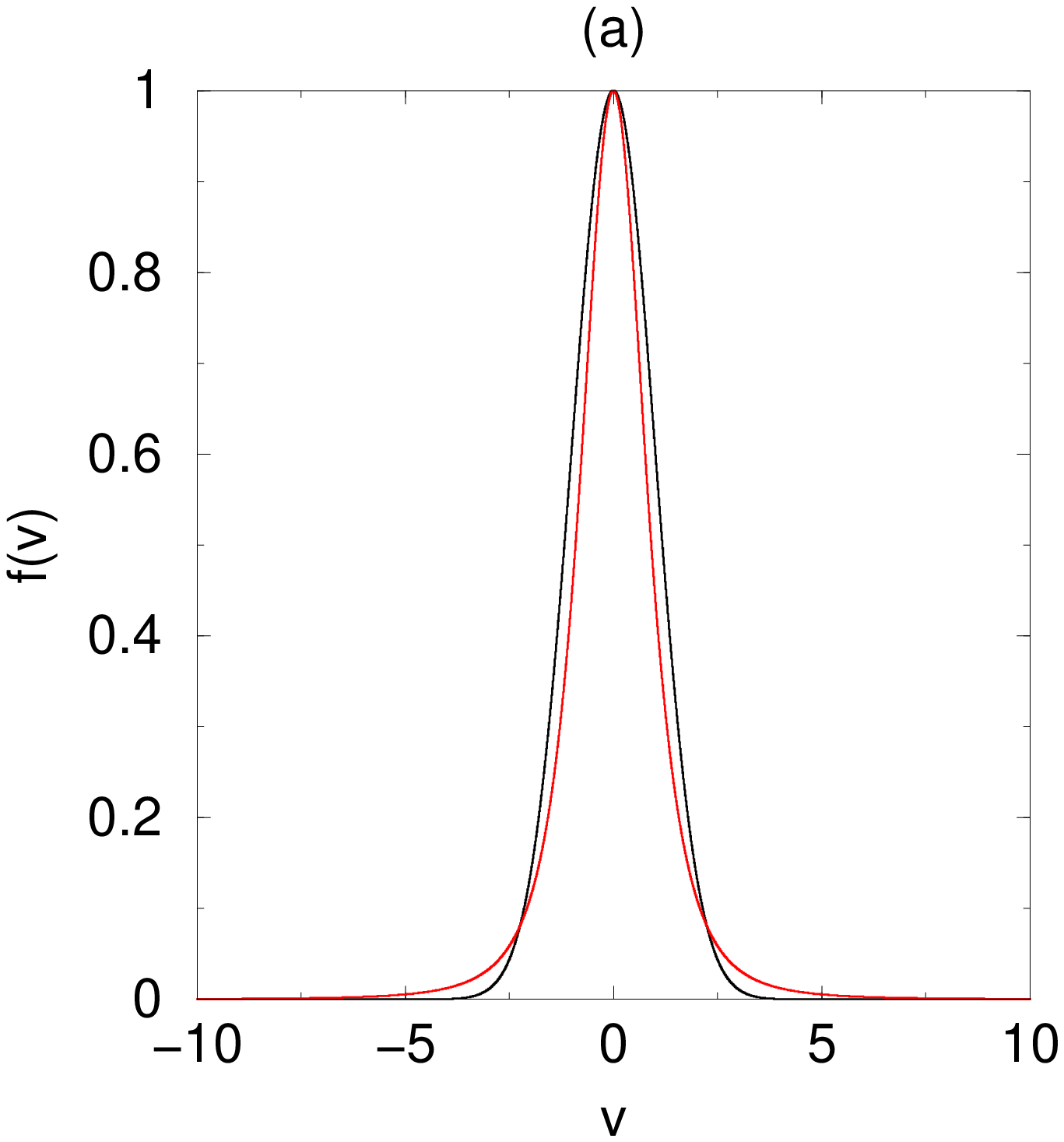}
  \includegraphics[height=10cm]{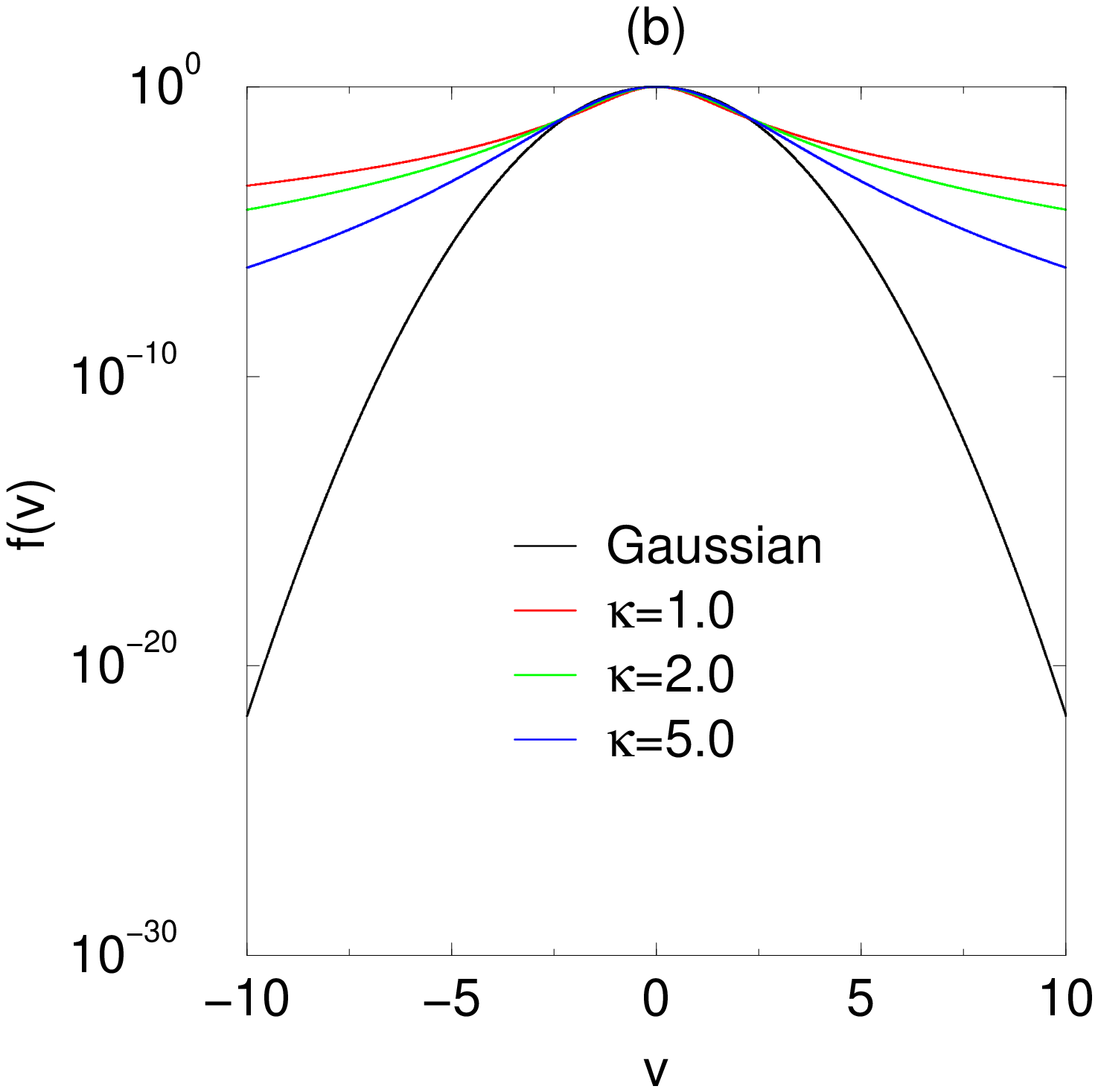}
\caption{An example of kappa distribution function:
(a) Maxwellian and $\kappa=1$ function compared in regular scale; 
(b) $\kappa=1$ (red) , $\kappa=2$ (green), $\kappa=5$ (blue), 
and Maxwellian (black) compared in a logarithmic scale.}
\label{fig4}
\end{figure}

\newpage
\begin{figure}[ht]
  \includegraphics[height=10cm] {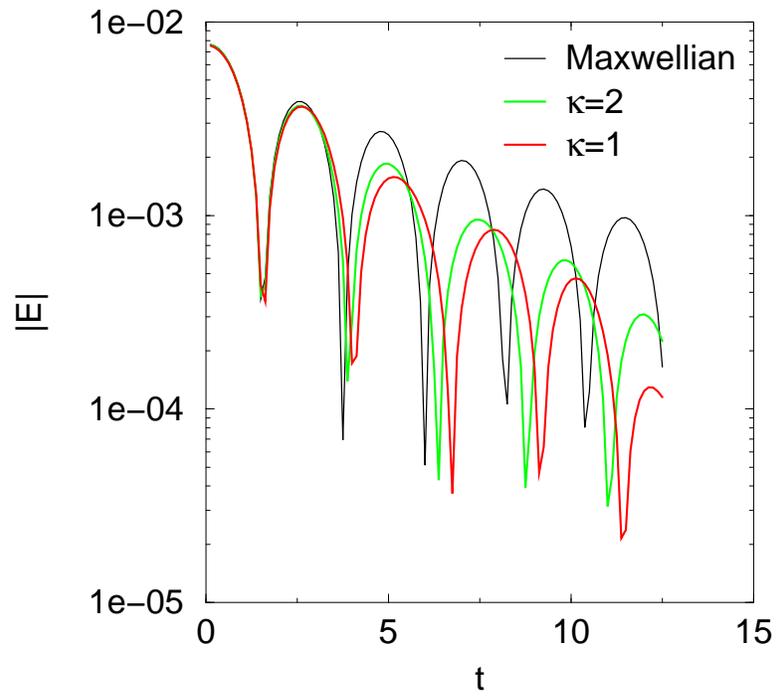}
\caption{Linear damping of electric field strength $|E|$ comparing
the cases when Maxwellian (black) and $\kappa=1$ (red) and $\kappa=2$
distribution functions are taken as initial conditions.}
\label{fig5}
\end{figure}

\newpage
\begin{figure}[ht]
  \includegraphics[height=10cm]{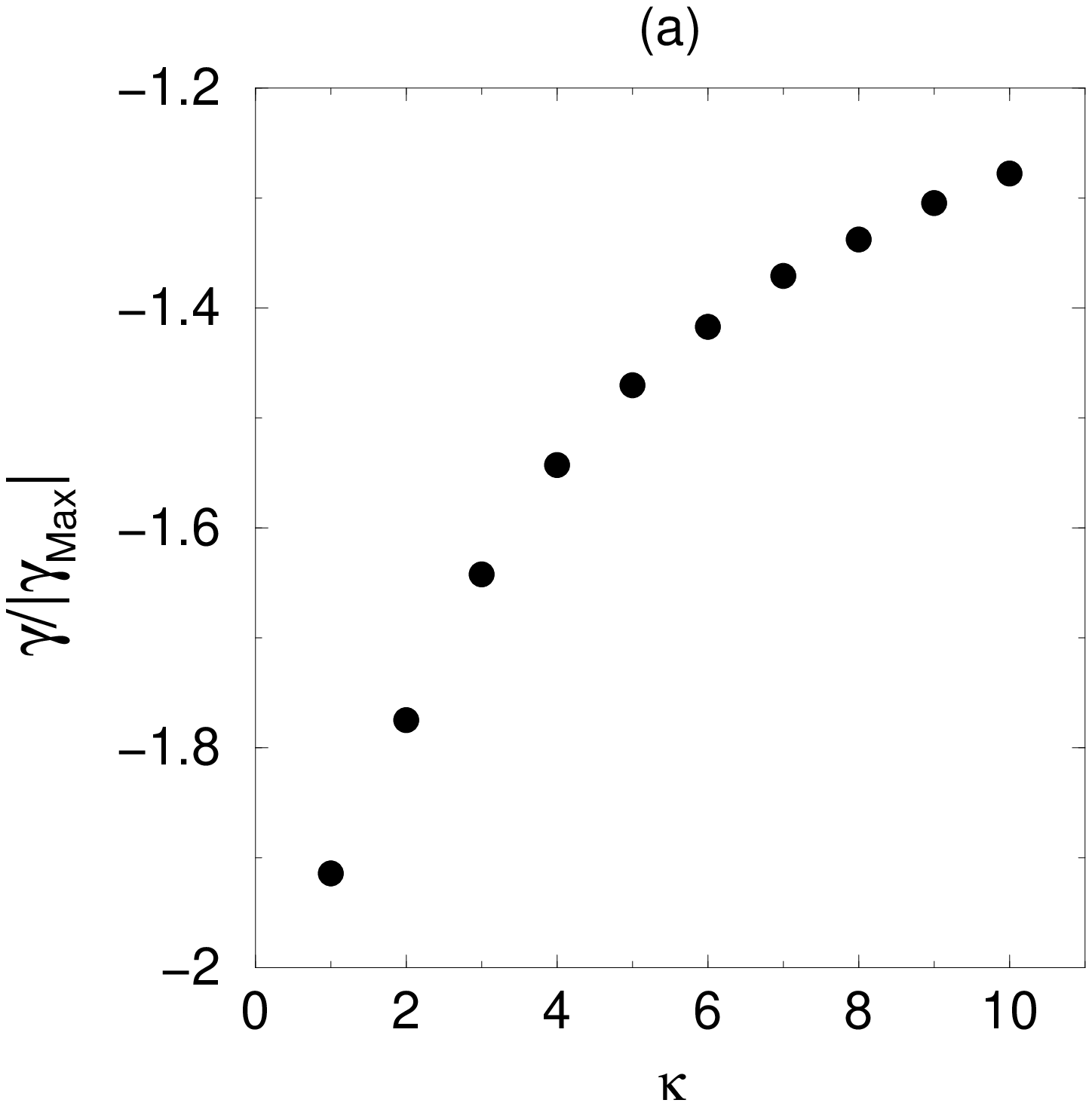}
  \includegraphics[height=10cm]{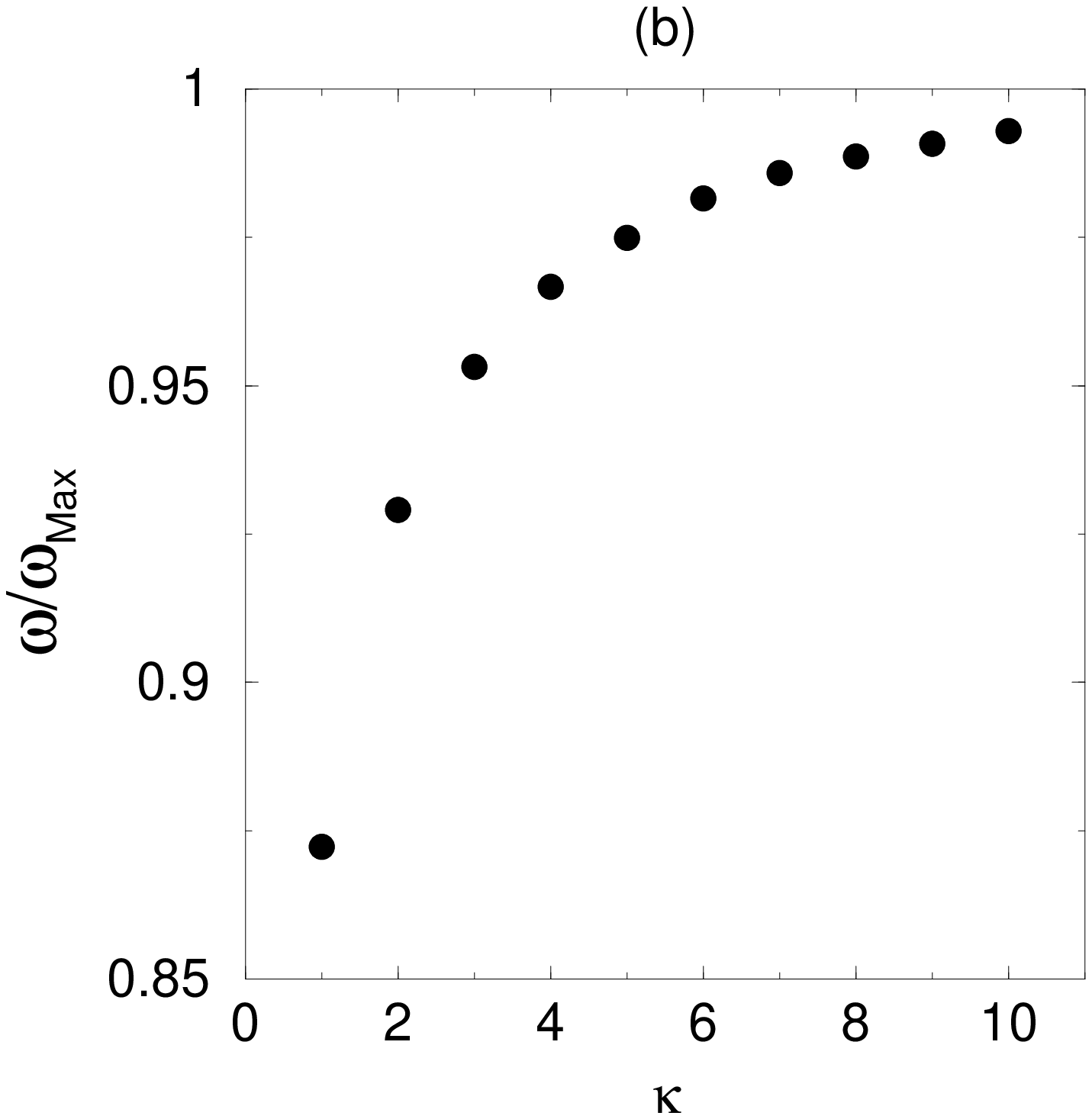}
\caption{Effect of kappa function summarized: (a) Damping rate
normalized by the absolute value of damping rate with Maxwellian; (b) real frequency
normalized by the  value of frequency with Maxwellian.}
\label{fig6}
\end{figure}

\newpage
\begin{figure}[ht]
  \includegraphics[height=10cm]{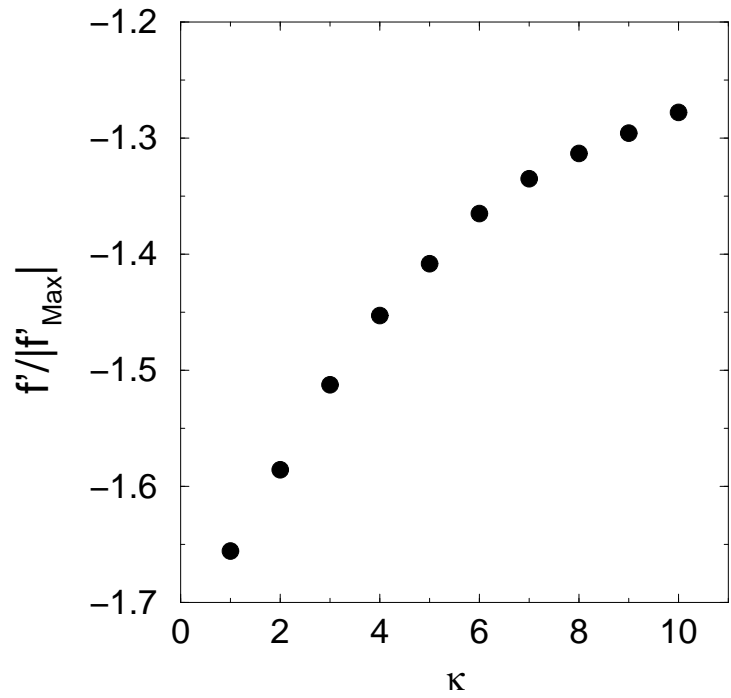}
\caption{The absolute value of $\partial_v f(v)$ at the phase velocity.
For the phase velocity $\omega /k$, the values of $\omega$ are
employed from Fig.6(b).
The plotted values are normalized by that of the Maxwellian.}
\label{fig7}
\end{figure}

\newpage
\begin{figure}[ht]
  \includegraphics[height=8cm]{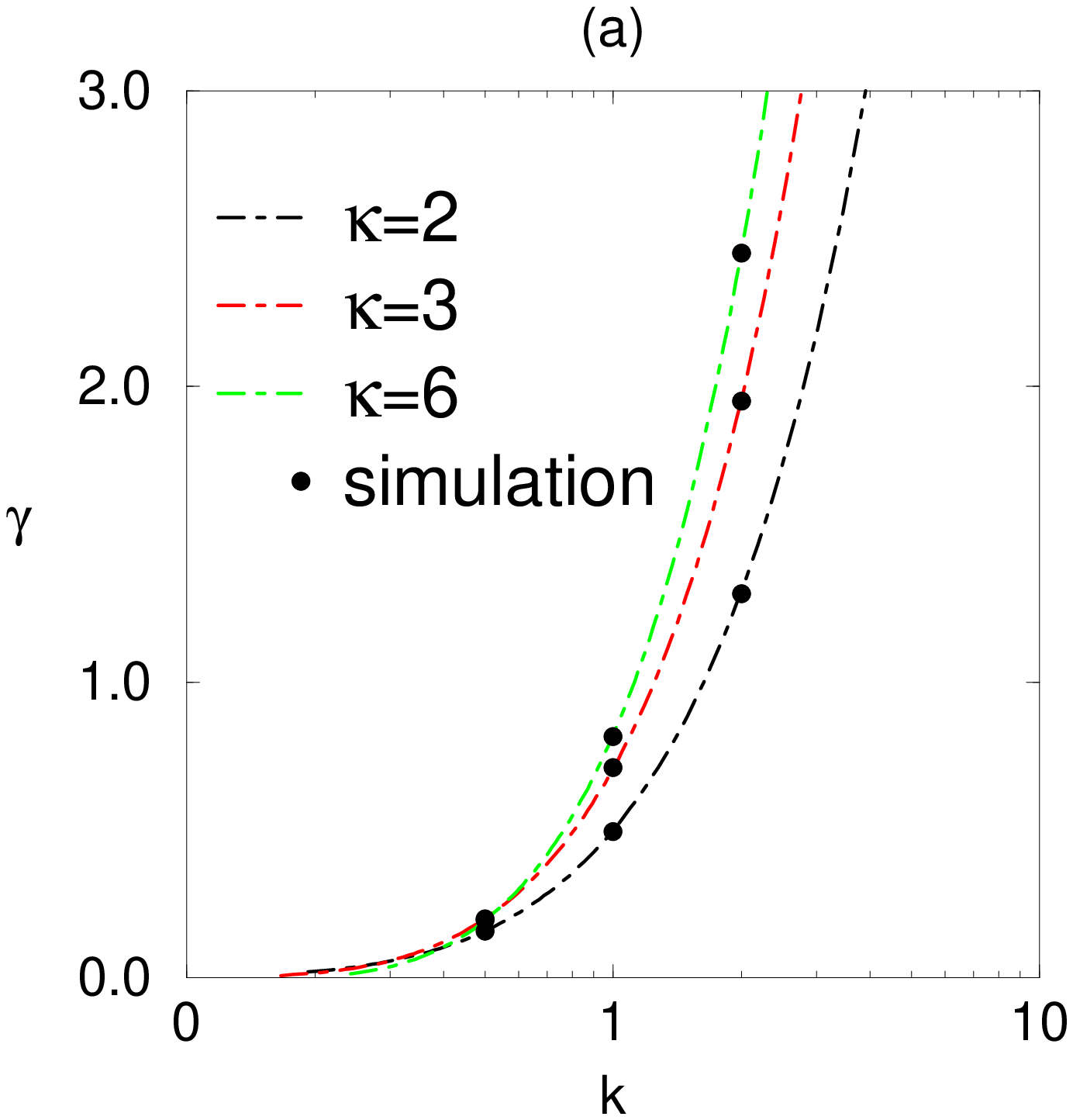}
  \includegraphics[height=8cm]{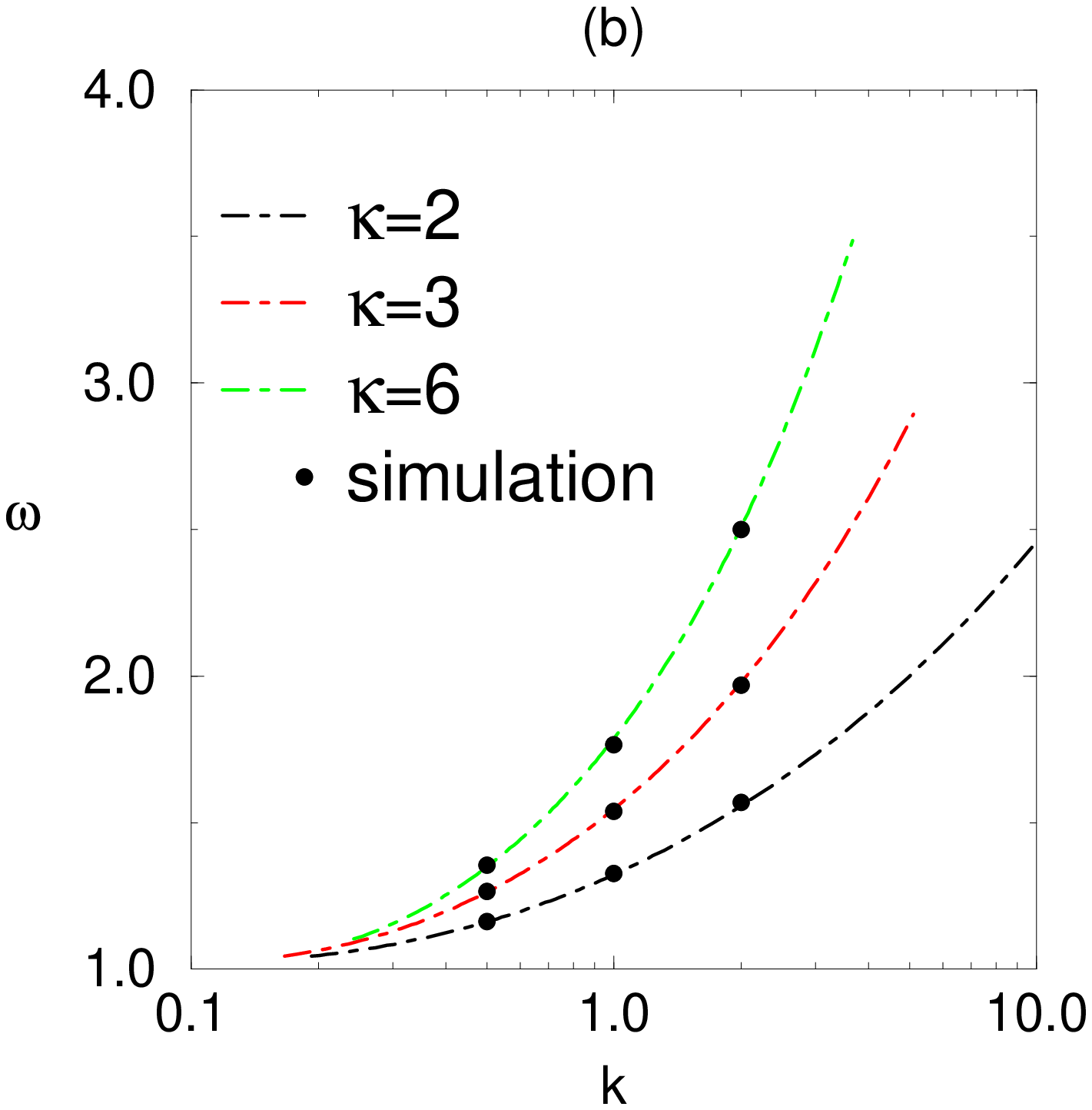}
\caption{Comparison of the numerical roots of the dispersion relation
and the simulation results:
(a) The damping rates $\gamma$ versus the wave vector $k$  
;(b) the real frequencies $\omega$ versus $k$. 
The roots from the (modified) dispersion relation are plotted as dash-dotted curves. 
The black, red, and green dash-dotted curves are for
$\kappa=2,3$, and $6$, respectively.
Black circles are obtained from linear numerical simulation. 
Numerical simulation is done for $k=0.5, 1.0$, and $2.0$ for $\kappa=2,3$, and $6$.
The simulation results are given at a fixed point $x=\pi$.
}
\label{fig8}
\end{figure}

\newpage
\begin{figure}[ht]
  \includegraphics[height=7cm]{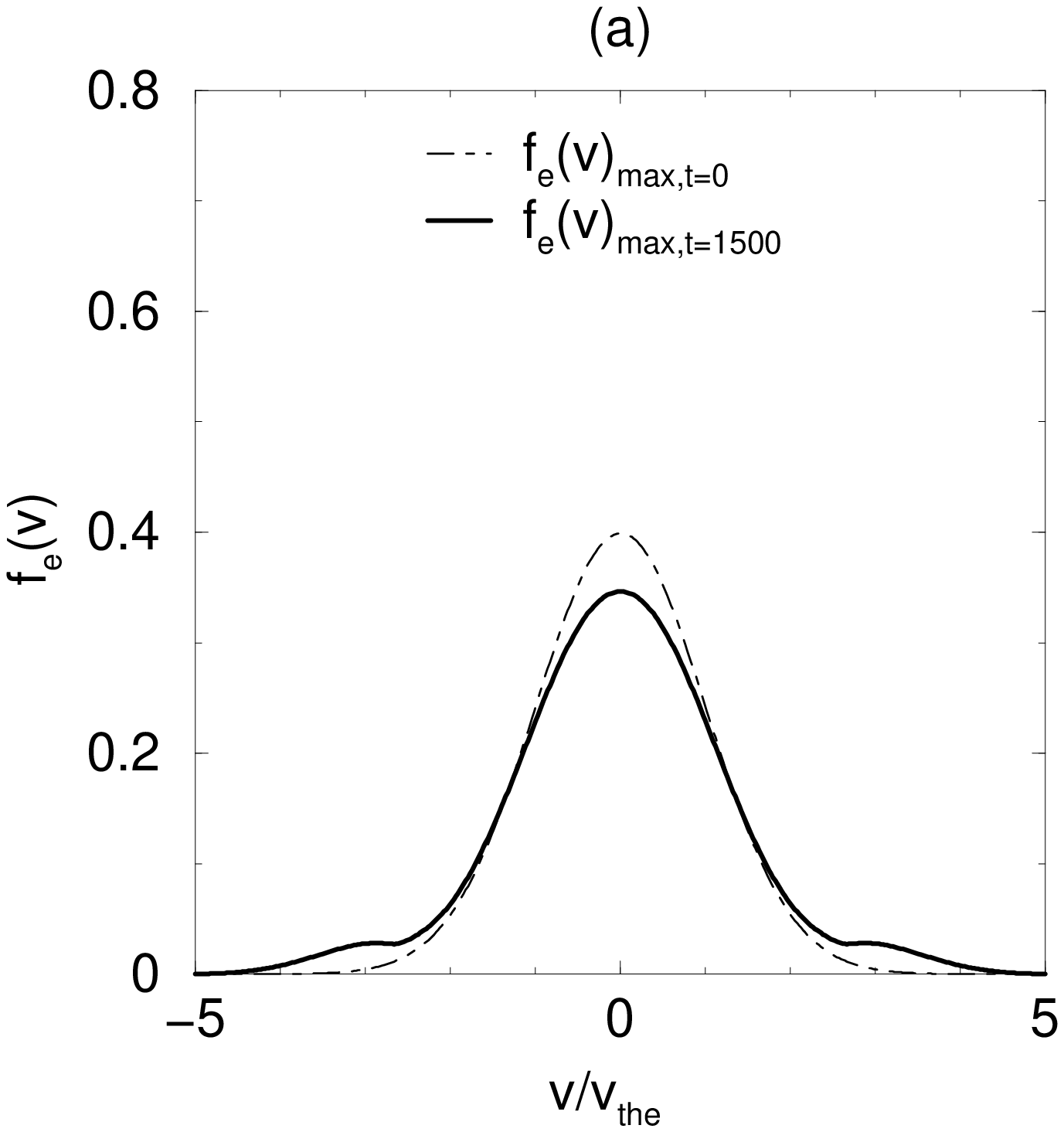}
  \includegraphics[height=7cm]{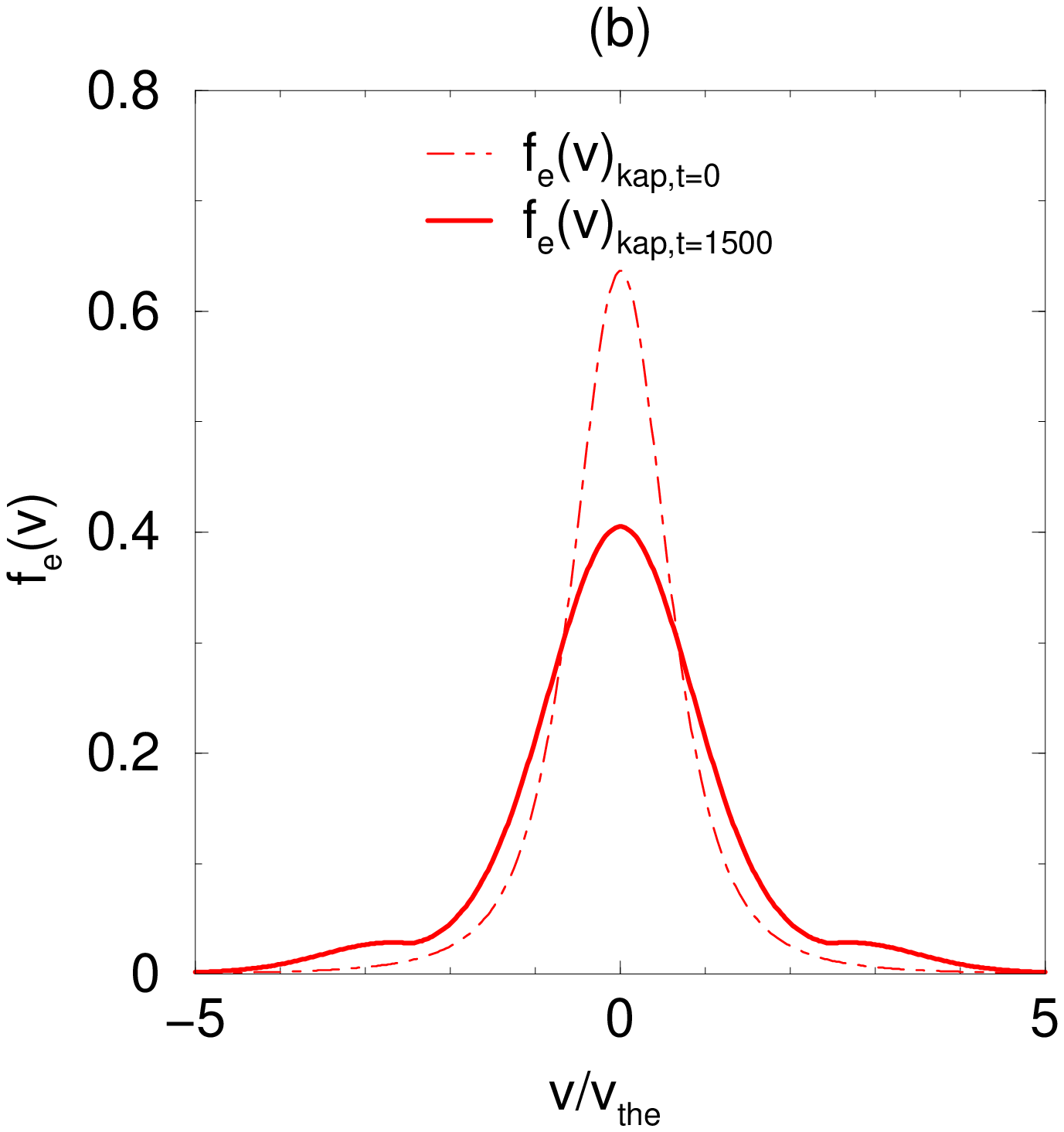}
  \includegraphics[height=7cm]{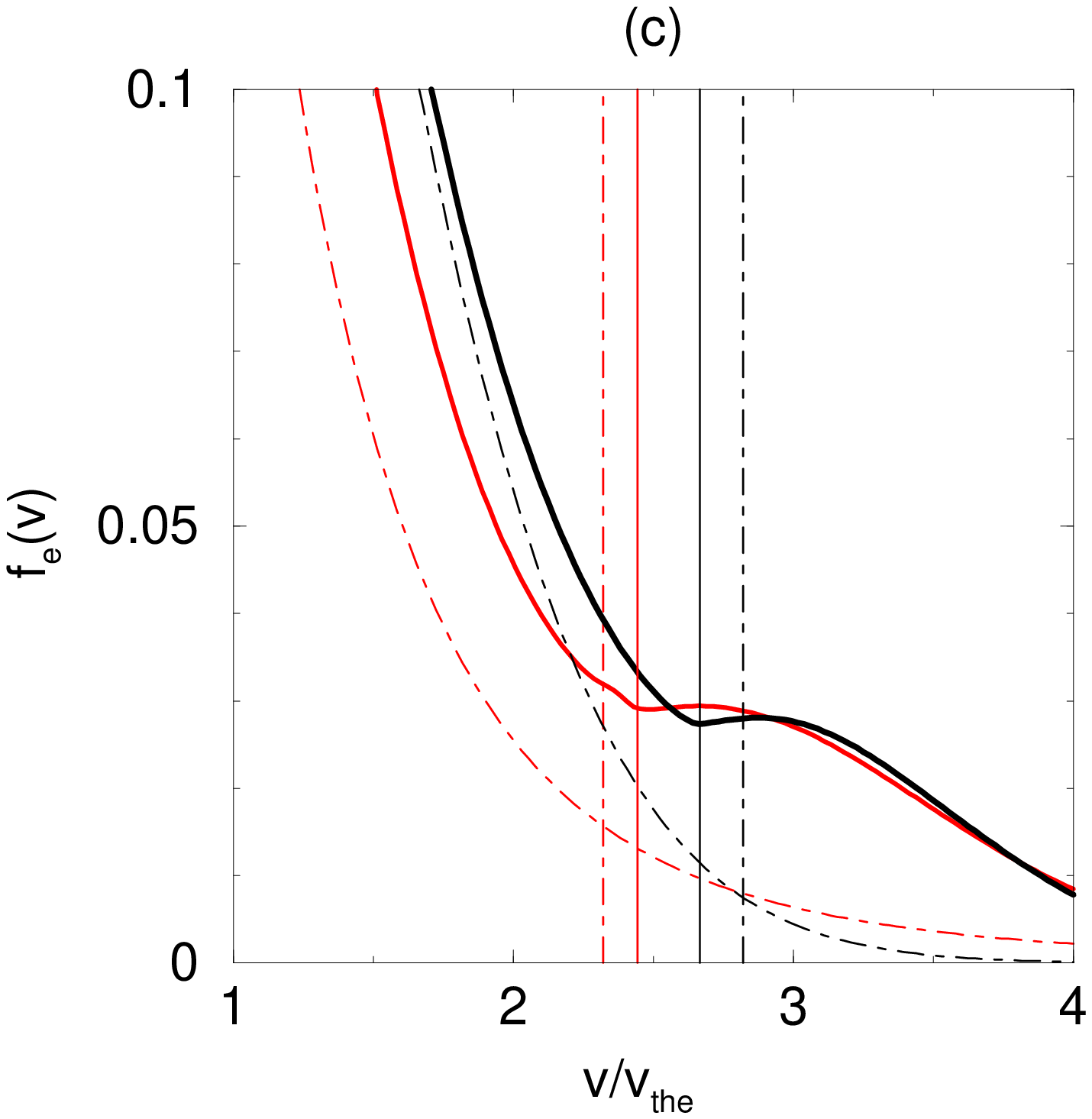}
\caption{Electron distribution functions suggesting long time evolution up to $t=1500$.
The distribution functions are given at a fixed point $x=\pi$.
(a) For a Maxwellian,
(b) for a kappa distribution function ($\kappa=2$). 
The dash-dotted curves are for $t=0$ and the solid curves are for $t=1500$.
(c) Local expansion of (a) and (b) near the resonant phase velocities.
Here, the black dash-dotted line at $\omega/k=2.82$ is for Maxwellian at $t=0$,
the black solid line at $\omega/k=2.66$ is for Maxwellian at $t=1500$,
the red dash-dotted line at $\omega/k=2.32$ is for kappa function at $t=0$,
and the red solid line at $\omega/k=2.47$ is for kappa function at $t=1500$, respectively.
The subscripts in the figure, $max$ and $kap$ stand for
the Maxwellian and the kappa distribution function, respectively.}
\label{fig9}
\end{figure}

\end{document}